# A Novel Algorithm for Informative Meta Similarity Clusters Using Minimum Spanning Tree

S. John Peter
Assistant Professor
Department of Computer Science and Research Center
St. Xavier's College, Palayamkottai
Tamil Nadu, India.
jaypeeyes@rediffmail.com

S. P. Victor
Associate Professor
Department of Computer Science and Research Center
St. Xavier's College, Palayamkottai
Tamil Nadu, India.
victorsp@rediffmail.com

ABSTRACT - The minimum spanning tree clustering algorithm is capable of detecting clusters with irregular boundaries. In this paper we propose two minimum spanning trees based clustering algorithm. The first algorithm produces k clusters with center and guaranteed intra-cluster similarity. The radius and diameter of k clusters are computed to find the tightness of k clusters. The variance of the k clusters are also computed to find the compactness of the clusters. The second algorithm is proposed to create a dendrogram using the k clusters as objects with guaranteed inter-cluster similarity. The algorithm is also finds central cluster from the k number of clusters. The first algorithm uses divisive approach, where as the second algorithm uses agglomerative approach. In this paper we used both the approaches to find Informative Meta similarity clusters.

*Key Words: Euclidean minimum spanning tree, Subtree, Eccentricity, Center, Tightness. Hierarchical clustering, Dendrogram, Radius, Diameter, Central clusters, Compactness*.

## I INTRODUCTION

Given a connected, undirected graph $G=(V,E)$, where $V$ is the set of nodes, $E$ is the set of edges between pairs of nodes, and a weight $w(u, v)$ specifying weight of the edge $(u, v)$ for each edge $(u, v) \in E$. A spanning tree is an acyclic subgraph of a graph G, which contain all vertices from $G$. The Minimum Spanning Tree (**MST**) of a weighted graph is minimum weight spanning tree of that graph. Several well established **MST** algorithms exist to solve minimum spanning tree problem [21, 15, 17]. The cost of constructing a minimum spanning tree is $O(m \log n)$, where $m$ is the number of edges in the graph and $n$ is the number of vertices. More efficient algorithm for constructing **MST**s have also been extensively researched [13, 5, 8]. These algorithms promise close to linear time complexity under different assumptions. A Euclidean minimum spanning tree (**EMST**) is a spanning tree of a set of $n$ points in a metric space ($E^n$), where the length of an edge is the Euclidean distance between a pair of points in the point set.

The hierarchical clustering approaches are related to graph theoretic clustering. Clustering algorithms using minimal spanning tree takes the advantage of **MST**. The **MST** ignores many possible connections between the data patterns, so the cost of clustering can be decreased. The **MST** based clustering algorithm is known to be capable of detecting clusters with various shapes and size [24]. Unlike traditional clustering algorithms, the **MST** clustering algorithm does not assume a spherical shapes structure of the underlying data. The **EMST** clustering algorithm [20, 24] uses the Euclidean minimum spanning tree of a graph to produce the structure of point clusters in the $n$-dimensional Euclidean space. Clusters are detected to achieve some measures of optimality, such as minimum intra-cluster distance or maximum inter-cluster distance [2]. The **EMST** algorithm has been widely used in practice.

Clustering by minimal spanning tree can be viewed as a hierarchical clustering algorithm which follows a divisive approach. Using this method firstly **MST** is constructed for a given input. There are different methods to produce group of clusters. If the number of clusters $k$ is given in advance, the simplest way to obtain $k$ clusters is to sort the edges of minimum spanning tree in descending order of their weights and remove edges with first *k*-1 heaviest weights [2, 23].

Geometric notion of centrality are closely linked to facility location problem. The distance matrix *D* can computed rather efficiently using Dijkstra's algorithm with time complexity $O(|V|^2 \ln |V|)$ [22].

The *eccentricity* of a vertex *x* in *G* and radius $\rho(G)$, respectively are defined as
$$e(x) = \max_{y \in V} d(x, y) \quad \text{and} \quad \rho(G) = \min_{x \in V} e(x)$$

The *center* of *G* is the set
$$C(G) = \{x \in V \mid e(x) = \rho(G)\}$$

*C* (G) is the center to the "*emergency facility location problem*" which is always contain single block of *G*. The length of the longest path in the graph is called *diameter* of the graph *G*. we can define diameter D (G) as
$$D(G) = \max_{x \in V} e(x)$$

The *diameter* set of *G* is
$$Dia(G) = \{x \in V \mid e(x) = D(G)\}$$

All existing clustering Algorithm require a number of parameters as their inputs and these parameters can significantly affect the cluster quality. In this paper we want to avoid experimental methods and advocate the idea of need-specific as opposed to care-specific because users always know the needs of their applications. We believe it is a good idea to allow users to define their desired similarity within a cluster and allow them to have some flexibility to adjust the similarity if the adjustment is needed. Our Algorithm produces clusters of *n*-dimensional points with a given cluster number and a naturally approximate intra-cluster distance.

Hierarchical clustering is a sequence of partitions in which each partition is nested into the next in sequence. An Agglomerative algorithm for hierarchical clustering starts with disjoint clustering, which places each of the *n* objects in an individual cluster [1]. The hierarchical clustering algorithm being employed dictates how the proximity matrix or proximity graph should be interpreted to merge two or more of these trivial clusters, thus nesting the trivial clusters into second partition. The process is repeated to form a sequence of nested clustering in which the number of clusters decreases as a sequence progress until single cluster containing all *n* objects, called the *conjoint clustering*, remains[1].

An important objective of hierarchical cluster analysis is to provide picture of data that can easily be interpreted. A picture of a hierarchical clustering is much easier for a human being to comprehend than is a list of abstract symbols. A *dendrogram* is a special type of tree structure that provides a convenient way to represent hierarchical clustering. A dendrogram consists of layers of nodes, each representing a cluster.

In this paper we propose two **EMST** based clustering algorithm to address the issues of undesired clustering structure and unnecessary large number of clusters. Our first algorithm assumes the number of clusters is given. The algorithm constructs an **EMST** of a point set and removes the inconsistent edges that satisfy the inconsistence measure. The process is repeated to create a hierarchy of clusters until *k* clusters are obtained. In section 2 we review some of the existing works on graph based clustering algorithm and central tree in a Minimum Spanning Trees. In Section 3 we propose **EMSTRD** algorithm which produces *k* clusters with center, radius, diameter and variance. We also propose another algorithm called **EMSTUCC** for finding cluster of clusters using the *k* clusters which are from previous **EMSTRD** algorithm. The algorithm also finds central cluster. Hence we named this new cluster as *Informative Meta similarity clusters*. The radius, diameter and variance of sub tree(cluster) is used to find tightness and compactness of clusters. Finally in conclusion we summarize the strength of our methods and possible improvements.

## II. RELATED WORK

Clustering by minimal spanning tree can be viewed as a hierarchical clustering algorithm which follows the divisive approach. Clustering Algorithm based on minimum and maximum spanning tree were extensively studied. Avis [3] found an $O(n^2 \log^2 n)$ algorithm for the min-max diameter-2 clustering problem. Asano, Bhattacharya, Keil and Yao [2] later gave optimal $O(n \log n)$ algorithm using maximum spanning trees for minimizing the maximum diameter of a bipartition. The problem becomes NP-complete when the number of partitions is beyond two [12]. Asano, Bhattacharya, Keil and Yao also considered the clustering problem in which the goal to maximize the minimum inter-cluster distance. They gave a *k*-partition of point set removing the *k*-1 longest edges from the minimum spanning tree constructed from that point set [2]. The identification of inconsistent edges causes problem in the **MST** clustering algorithm. There exist numerous ways to divide clusters successively, but there is not suitable choice for all cases.

Zahn [24] proposes to construct **MST** of point set and delete inconsistent edges – the edges, whose weights are significantly larger than the average weight of the nearby edges in the tree. Zahn's inconsistent measure is defined as follows. Let $e$ denote an edge in the **MST** of the point set, $v_1$ and $v_2$ be the end nodes of $e$, $w$ be the weight of $e$. A *depth neighborhood N* of an end node $v$ of an edge $e$ defined as a set of all edges that belong to all the path of length $d$ originating from $v$, excluding the path that include the edge $e$. Let $N_1$ and $N_2$ be the depth $d$ neighborhood of the node $v_1$ and $v_2$. Let $\hat{W}_{N1}$ be the average weight of edges in $N_1$ and $\sigma N_1$ be its standard deviation. Similarly, let $\hat{W}_{N2}$ be the average weight of edges in $N_2$ and $\sigma N_2$ be its standard deviation. The inconsistency measure requires one of the three conditions hold:

1. $w > \hat{W}N_1 + c \times \sigma N_1$ or $w > \hat{W}N_2 + c \times \sigma N_2$

2. $w > max(\hat{W}N_1 + c \times \sigma N_1, \hat{W}N_2 + c \times \sigma N_2)$

3. $\dfrac{w}{max(c \times \sigma N_1, c \times \sigma N_2)} > f$

where $c$ and $f$ are preset constants. All the edges of a tree that satisfy the inconsistency measure are considered inconsistent and are removed from the tree. This result in set of disjoint subtrees each represents a separate cluster. Paivinen [19] proposed a Scale Free Minimum Spanning Tree (**SFMST**) clustering algorithm which constructs scale free networks and outputs clusters containing highly connected vertices and those connected to them.

The **MST** clustering algorithm has been widely used in practice. Xu (Ying), Olman and Xu (Dong) [23] use MST as multidimensional gene expression data. They point out that **MST**- based clustering algorithm does not assume that data points are grouped around centers or separated by regular geometric curve. Thus the shape of the cluster boundary has little impact on the performance of the algorithm. They described three objective functions and the corresponding cluster algorithm for computing $k$-partition of spanning tree for predefined $k > 0$. The algorithm simply removes $k$-1 longest edges so that the weight of the subtrees is minimized. The second objective function is defined to minimize the total distance between the center and each data point in the cluster. The algorithm removes first $k$-1 edges from the tree, which creates a $k$-partitions.

The clustering algorithm proposed by S.C.Johnson [11] uses proximity matrix as input data. The algorithm is an agglomerative scheme that erases rows and columns in the proximity matrix as old clusters are merged into new ones. The algorithm is simplified by assuming no ties in the proximity matrix. Graph based algorithm was proposed by Hubert [7] using single link and complete link methods. He used threshold graph for formation of hierarchical clustering. An algorithm for single-link hierarchical clustering begins with the minimum spanning tree (MST) for G ($\infty$), which is a proximity graph containing $n(n-1)/2$ edge was proposed by Gower and Ross [9]. Later Hansen and DeLattre [6] proposed another hierarchical algorithm from graph coloring.

Given $n$ d-dimensional data objects or points in a cluster, we can define the centroid $x_0$, radius $R$, diameter $D$ and variance of the cluster as

$$X_0 = \dfrac{\sum_{i=1}^{n} X_i}{n}$$

$$R = \left( \dfrac{\sum_{i=1}^{n} (X_i - X_0)^2}{n} \right)^{1/2}$$

$$D = \left( \dfrac{\sum_{i=1}^{n}\sum_{j=1}^{n} (X_i - X_j)^2}{n(n-1)} \right)^{1/2}$$

where $R$ is the average distance from member objects to the centroid, and $D$ is the average pairwise distance within a cluster. Both $R$ and $D$ reflect the tightness of the cluster around centroid[25].

The cospanning tree of a tree spanning tree $T$ is edge complement of $T$ in $G$. Also the rank $\rho(G)$ of a Graph $G$ with $n$ vertices and $k$ connected components is $n$-$k$.
A central tree[4] of a graph is a tree $T_0$ such that the rank $r$ of its cospanning tree $T_o$ is minimum.

$$r = \rho\ (T_0) \leq \rho\ (T),\ \forall T \in G.$$

Deo[4] pointed out that, if $r$ is the rank of the cospanning tree of $T$, then there is no tree in $G$ at a distance greater than $r$ from $T$ and there is at least one tree in $G$ at distance exactly $r$ from $T$. A direct consequence of this is following characterization of central tree.

A Spanning tree $T_0$ is a central tree of $G$ if and only if the largest distance from $T_0$ to any other tree in $G$ is minimum[4], ie,

$max\ d(T_0,T_i) \leq max\ d(T,T_i), \forall\ T_i \in G$

The *maximally distant* tree problem, for instance, which ask for a pair of spanning tree($T_1,T_2$) such that $d(T_1,T_2) \geq d(T_i,T_j), \forall\ T_i,T_j \in G$, can be solved in polynomial time[14]. Also, as pointed out in [16], the distance between tree pairs a graph $G$ are in a one-to-one correspondence with the distance between vertex pairs in the *tree-graph G*. Thus finding a central tree in $G$ is equivalent to finding a central vertex in a tree of $G$. However, while central vertex problem is known to have a polynomial time algorithm (in number of vertices), such an algorithm can not be used efficiently find a central tree, since the number of vertices in a tree of $G$ can be exponential.

### III. OUR CLUSTERING ALGORITHM

A tree is a simple structure for representing binary relationship, and any connected components of tree is called *subtree*. Through this **MST** representation, we can convert a multi-dimensional clustering problem to a tree partitioning problem, i.e., finding particular set of tree edges and then cutting them. Representing a set of multi-dimensional data points as simple tree structure will clearly lose some of the inter data relationship. However many clustering algorithm proved that no essential information is lost for the purpose of clustering. This is achieved through rigorous proof that each cluster corresponds to one subtree, which does not overlap the representing subtree of any other cluster. Clustering problem is equivalent to a problem of identifying these subtrees through solving a tree partitioning problem. The inherent cluster structure of a point set in a metric space is closely related to how objects or concepts are embedded in the point set. In practice, the approximate number of embedded objects can sometimes be acquired with the help of domain experts. Other times this information is hidden and unavailable to the clustering algorithm. In this section we preset clustering algorithm which produce $k$ clusters, with center, radius, diameter and variance of each cluster. We also present another algorithm to find the hierarchy of $k$ clusters and central cluster.

#### A. Cluster Tightness Measure and Compactness

In order to measure the efficacy of clustering, a measure based upon the radius and diameter of each subtree (cluster) is devised. The radius and diameter values of each cluster are expected low value for good cluster. If the values are large that the points (objects) are spread widely and may overlap. The cluster tightness measure is a within – cluster estimate of clustering effectiveness, however it is possible to devise inter- cluster measure also, to better measure the separation between the various clusters.

The Cluster compactness measure is based on the variance of the data points distributed in the subtrees (clusters). The variance of cluster $T$ is computed as

$$v(T) = \left( \frac{1}{n} \sum_{i=1}^{n} d^2(x_i, x_0) \right)^{1/2}$$

Where $d(x_i, x_j)$ is distance metric between two points(objects) $x_i$ and $x_j$, where n is the number of objects in the subtree $T_i$, and $x_0$ is the mean of the subtree $T$. A smaller the variance value indicates, a higher homogeneity of the objects in the data set, in terms of the distance measure $d(\ )$. Since $d(\ )$ is the Euclidean distance, $v(T_i)$ becomes the statistical variance of data set $\sigma(T_i)$. The cluster compactness for the out put clusters generated by the algorithm is the defined as

$$Cmp = \frac{1}{k} \sum_{i=1}^{k} \frac{v(T_i)}{v(S)}$$

Where $k$ is the number of clusters generated on the given data set $S$, $v(T_i)$ is the variance of the clusters $T_i$ and $V(S)$ is the variance of data set $S$. [10]

The cluster compactness measure evaluates how well the subtrees (clusters) of the input is redistributed in the clustering process, compared with the whole input set, in terms of data homogeneity reflected by Euclidean distance metric used by the clustering process. Smaller the cluster compactness value indicates a higher average compactness in the out put clusters.

*B. EMSTRD Clustering Algorithm*

Given a point set $S$ in $\mathbf{E}^n$ and the desired number of clusters $k$, the hierarchical method starts by constructing an **MST** from the points in $S$. The weight of the edge in the tree is Euclidean distance between the two end points. Next the average weight $\hat{W}$ of the edges in the entire **EMST** and its standard deviation $\sigma$ are computed; any edge with ($W > \hat{W} + \sigma$) or (*current longest edge*) is removed from the tree. This leads to a set of disjoint subtrees $S_T = \{T_1, T_2 \ldots\}$ *(divisive approach)*. Each of these subtrees $T_i$ is treated as cluster. Oleksandr Grygorash et al proposed algorithm [18] which generates $k$ clusters. We modified the algorithm in order to generate $k$ clusters with centers. The algorithm is also find radius, diameter and variance of subtrees, which is useful in finding tightness and compactness of clusters. Hence we named the new algorithm as Euclidean Minimum Spanning Tree for Radius and Diameter (**EMSTRD**). Each center point of $k$ clusters is a representative point for the each subtree $S_T$. A point $c_i$ is assigned to a cluster $i$ if $c_i \in T_i$. The group of center points is represented as $C = \{c_1, c_2 \ldots c_k\}$

**Algorithm:** EMSTRD($k$)

Input : $S$ the point set
Output : $k$ number of clusters with C (set of center points)

Let $e$ be an edge in the **EMST** constructed from $S$
Let $W_e$ be the weight of $e$
Let $\sigma$ be the standard deviation of the edge weights
Let $S_T$ be the set of disjoint subtrees of the **EMST**
Let $n_c$ be the number of clusters
1. Construct an **EMST** from $S$
2. Compute the average weight of $\hat{W}$ of all the edges
3. Compute standard deviation $\sigma$ of the edges
4. Compute variance of the set S
5. $S_T = \emptyset$; $n_c = 1$; C= $\emptyset$;
6. **Repeat**
7. **For** each $e \in$ **EMST**
8. If ($W_e > \hat{W} + \sigma$) or (current longest edge $e$)
9. Remove $e$ from **EMST**
10. $S_T = S_T \cup \{T'\}$ // $T''$ is new disjoint subtree
11. $n_c = n_c + 1$
12. Compute the center $C_i$ of $T_i$ using eccentricity of points
13. Compute the diameter of $T_i$ using eccentricity of points
14. Compute the variance of $T_i$
15. C = $U_{T_i} \in S_T \{C_i\}$
16 Until $n_c = k$
17. **Return** $k$ clusters with $C$

Euclidean Minimum Spanning Tree is constructed at line 1. Average of edge weights and standard deviation are computed at lines 2-3. The variance of input data set S is computed at line 4. Using the average weight and standard deviation, the inconsistent edge is identified and removed from Euclidean Minimum Spanning Tree (**EMST**) at lines (8-9). Subtree (cluster) is created at line 10. Radius, diameter and variance of subtree are computed at lines 12-14. Lines 9-15 in the algorithm are repeated until $k$ number of subtrees (clusters) are produced. The radius and diameter are good measure to find the tightness of clusters. The radius and diameter values of each cluster are expected low value for good cluster. If the values are large that the points (objects) are spread widely. However if the value of $k$ increases the radius and diameter decreases.

The variance for each subtree (cluster) is computed to find the compactness of clusters. A smaller the variance value indicates, a higher homogeneity of the objects in the data set. The cluster compactness measure evaluates how well the subtrees (clusters) of the input is redistributed in the clustering process, compared with the whole input set, in terms of data homogeneity reflected by Euclidean distance metric used by the clustering process. Smaller the cluster compactness value indicates a higher average compactness in the out put clusters.

Figure 1 illustrate a typical example of cases in which simply remove the $k$-1 longest edges does not necessarily output the desired cluster structure. Our algorithm finds the center, radius, diameter and variance of the each cluster which will be useful in many applications. Our algorithm will find 7 cluster structures ($k$=7). Figure 2 shows the possible distribution of the points in the two cluster structures with their radius; diameter and also theirs center points 5 and 3. Figure 3 shows a graph which shows the relationship between radius and diameter with subtrees (clusters). Lower the radius and diameter value means higher the tightness. The compactness of the subtrees is shown in the figure 4. Lower the values of variance means higher the homogeneity of the objects in subtree (cluster).

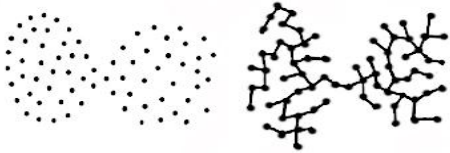

Figure 1. Clusters connected through a point

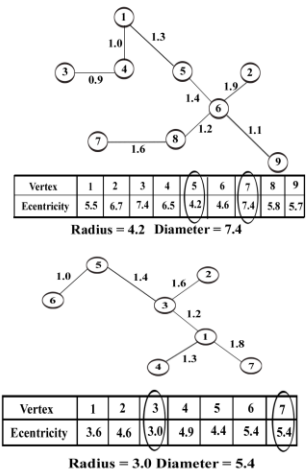

Figure 2. Two Clusters with radius and diameter (5 and 3 as center point)

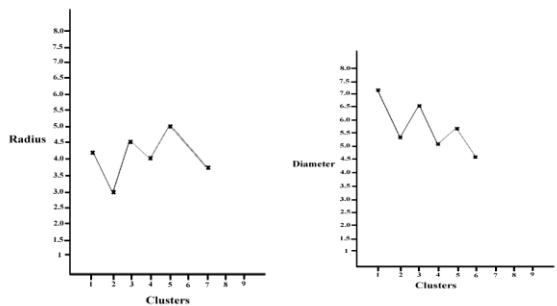

Figure 3: Tightness of clusters using radius and diameter

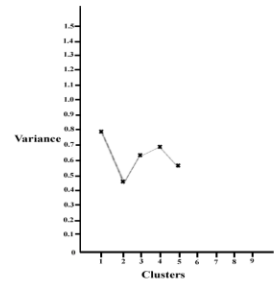

Figure 4. Compactness of clusters

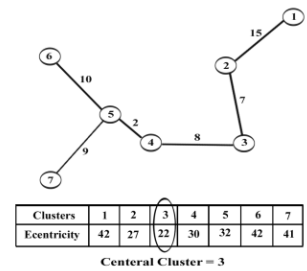

Figure 5. EMST From 7 cluster center points with central cluster 3

*C. EMSTUCC Algorithm for Central cluster*

The distance between two sub trees (clusters) of an **EMST** $T$ is defined as the number of edges present in one sub tree (cluster) but not present in the other.

$$d\ (T_1,\ T_2) = |T_1 - T_2| = |T_2 - T_1|$$

*Definition 1*: A sub tree (cluster) is a tree $T_0$ is a central sub tree (central cluster) of **EMST** $T$ if and only if the largest distance from $T_0$ to any other sub tree (cluster) in the **EMST** $T$ is minimum.

The result of the **EMSTRD** algorithm consists of $k$ number clusters with their center, radius, diameter and variance. These center points $c_1$, $c_2$ ....$c_k$ are connected and again minimum spanning tree is constructed is shown in the Figure 5. A Euclidean distance between pair of clusters can be represented by a corresponding weighted edge. Our Algorithm is also based on the minimum spanning tree but not limited to two-dimensional points. There were two kinds of clustering problem; one that minimizes the maximum intra-cluster distance and the other maximizes the minimum inter-cluster distances. Our Algorithms produces clusters with both intra-cluster and inter-cluster similarity. We propose Euclidean Minimum Spanning Tree Updation Central Cluster algorithm (**EMSTUCC**) converts the minimum spanning tree into dendrogram, which can be used to interpret about inter-cluster distances. This new algorithm is also finds central cluster from set of clusters.

The algorithm is neither single link clustering algorithm (SLCA) nor complete link clustering algorithm (CLCA) type of hierarchical clustering, but it is based on the distance between centers of clusters. This approach leads to new developments in hierarchical clustering. The level function, *L,* records the proximity at which each clustering is formed. The levels in the dendrogram tell us the least amount of similarity that points between clusters differ. This piece of information can be very useful in several medical and image processing applications.

**Algorithm:** EMSTUCC

Input: the point set C = {$c_1, c_2, \ldots, c_k$}
Output: central cluster and dendrogram

1. Construct an **EMST** *T* from *C*
2. Compute the radius of *T* using eccentricity of points // for central cluster
3. Begin with disjoint clusters with level $L(0) = 0$ and sequence number $m = 0$
4. **If** (*T* has some edge)
5. $e$ = get-min-edge(*T*) // for least dissimilar pair of clusters
6. $(i, j)$ = get-vertices ($e$)
7. Increment the sequence number $m = m + 1$, merge the clusters (*i*) and (*j*), into single cluster to form next clustering *m* and set the level of this cluster to $L(m) = e$;
8. Update *T* by forming new vertex by combining the vertices *i, j*; go to step 4.
9. **Else** Stop.

We use the graph of Figure 5 as example to illustrate the **EMSTUCC** algorithm. The **EMSTUCC** algorithm construct minimum spanning tree *T* from the points $c_1, c_2, c_3 \ldots c_k$ (line 1) and convert the *T* into dendrogram is shown in figure 6. The radius of the tree *T* is computed at line 2. This radius value is useful in finding the central cluster. It places the entire disjoint cluster at level 0 (line 3). It then checks to see if *T* still contains some edge (line 4). If so, it finds minimum edge *e* (line 5). It then finds the vertices i, j of an edge *e* (line 6). It then merges the vertices (clusters) and forms a new vertex (*agglomerative approach*). At the same time the sequence number is increased by one and the level of the new cluster is set to the edge weight (line 7). Finally, the Updation of minimum spanning tree is performed at line 8. The lines 4-8 in the algorithm are repeated until the minimum spanning tree *T* has no edge. The algorithm takes $O(|E|^2)$ time.

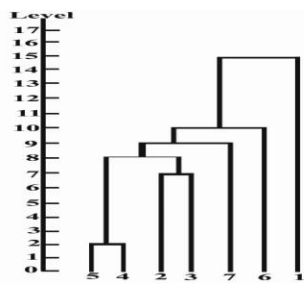

Figure 6. Dendrogram for Meta cluster

IV. CONCLUSION

Our **EMSTRD** clustering algorithm assumes a given cluster number. The algorithm gradually finds *k* clusters with center for each cluster. These *k* clusters ensures guaranteed intra-cluster similarity. The algorithm finds radius, diameter and variance of clusters using eccentricity of points in a cluster. The radius and diameter value gives the information about tightness of clusters. The variance value of the cluster is useful in finding the compactness of cluster. Our algorithm does not require the users to select and try various parameters combinations in order to get the desired output. All of these look nice from theoretical point of view. However from practical point of view, there is still some room for improvement for running time of the clustering algorithm. This could perhaps be accomplished by using some appropriate data structure. Our **EMSTUCC** clustering algorithm generates dendrogram which is used to find the relationship between the *k* number clusters produced from the **EMSTRD** algorithm. The inter-cluster distance between *k* clusters are shown in the Dendrogram.

The algorithm is also finds central cluster. This information will be very useful in many applications. In the future we will explore and test our proposed clustering algorithm in various domains. The **EMSTRD** algorithm uses divisive approach, where as the **EMSTUCC** algorithm uses agglomerative approach. In this paper we used both the approaches to find Informative Meta similarity clusters. We will further study the rich properties of **EMST**-based clustering methods in solving different clustering problems.

## REFERENCE

[1] Anil K. Jain, Richard C. Dubes *"Algorithm for Clustering Data" Michigan State University, Prentice Hall, Englewood Cliffs, New Jersey* 07632.1988.

[2] T.Asano, B.Bhattacharya, M.Keil and F.Yao. "Clustering Algorithms based on minimum and maximum spanning trees". In *Proceedings of the 4th Annual Symposium on Computational Geometry,* Pages 252-257, 1988.

[3] D.Avis "Diameter partitioning". *Discrete and Computational Geometry,* 1:265-276, 1986

[4] N.Deo, "A central tree", *IEEE Trans. Circuit theory CT-13* (1966) 439-440, correspondence.

[5] M.Fredman and D.Willard. "Trans-dichotomous algorithms for minimum spanning trees and shortest paths". In *Proceedings of the 31st Annual IEEE Symposium on Foundations of Computer Science,* pages 719-725, 1990.

[6] P. Hansen and M. Delattre " Complete-link cluster analysis by graph coloring" *Journal of the American Statistical Association* 73, 397-403, 1978.

[7] Hubert L. J " Min and max hierarchical clustering using asymmetric similarity measures " *Psychometrika* 38, 63-72, 1973.

[8] H.Gabow, T.Spencer and R.Rajan. " Efficient algorithms for finding minimum spanning trees in undirected and directed graphs". *Combinatorica,* 6(2):109-122, 1986.

[9] J.C. Gower and G.J.S. Ross "Minimum Spanning trees and single-linkage cluster analysis" *Applied Statistics* 18, 54-64, 1969.

[10] Ji He, Ah-Hwee Tan, Chew- Lim Tan, Sam- Yuan Sung. "On Quantitative Evaluation of clustering systems. Information Retrieval and Clustering". W.Wu and H.Xiong (Eds.). Kluwer Academic Publishers. 2002

[11] S. C. Johnson "Hierarchical clustering schemes" *Psychometrika* 32, 241-254, 1967.

[12] D,Johnson. "The np-completeness column: An ongoing guide". *Journal of Algorithms,*3:182-195, 1982.

[13] D.Karger, P.Klein and R.Tarjan. "A randomized linear-time algorithm to find minimum spanning trees". *Journal of the ACM,* 42(2):321-328, 1995.

[14] G.Kishi, Y.Kajitani, "Maximally distant trees and principal partition of linear graph". *IEEE Trans. Circuit Theory CT*-16(1969) 323-330.

[15] J.Kruskal. "On the shortest spanning subtree and the travelling salesman problem". *In Proceedings of the American Mathematical Society,* Pages 48-50, 1956.

[16] N.Malik, Y.Y.Lee, "Finding trees and singed tree pairs by the compound method", *in Tenth Midwest Symposium on Circuit Theory,* 1967, pp. V1-5-1-V1-5-11.

[17] J.Nesetril, E.Milkova and H.Nesetrilova. Otakar boruvka "on minimum spanning tree problem: Translation of both the 1926 papers, comments, history. DMATH*:" Discrete Mathematics,* 233, 2001.

[18] Oleksandr Grygorash, Yan Zhou, Zach Jorgensen. "Minimum spanning Tree Based Clustering Algorithms". *Proceedings of the 18th IEEE International conference on tools with Artificial Intelligence (ICTAI'06)* 2006.

[19] N.Paivinen. "Clustering with a minimum spanning of scale-free-like structure".*Pattern Recogn. Lett.,*26(7): 921-930, 2005.

[20] F.Preparata and M.Shamos. "Computational Geometry: An Introduction". *Springer-Verlag, Newyr, NY ,USA*, 1985

[21] R.Prim. "Shortest connection networks and some generalization". *Bell systems Technical Journal,*36:1389-1401, 1957.

[22] Stefan Wuchty and Peter F. Stadler. "Centers of Complex Networks". 2006

[23] Y.Xu, V.Olman and D.Xu. "Minimum spanning trees for gene expression data clustering". *Genome Informatics,*12:24-33, 2001.

[24] C.Zahn. "Graph-theoretical methods for detecting and describing gestalt clusters". *IEEE Transactions on Computers*, C-20:68-86, 1971.

[25] T.Zhang, R.Ramakrishnan and M.Livny. "BIRCH: an efficient data clustering method for very large databases". In *Proc.1996 ACM-SIGMOD Int. Conf. Management of Data ( SIGMOD'96),* pages 103-114, Montreal, Canada, June 1996.

## BIOGRAPHY OF AUTHORS

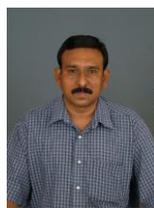

**S. John Peter is** working as Assistant professor in Computer Science, St.Xavier's college (Autonomous), Palayamkottai, Tirunelveli. He earned his M.Sc degree from Bharadhidasan University, Trichirappli. He also earned his M.Phil from Bhradhidasan University, Trichirappli. Now he is doing Ph.D in Computer Science at Manonmaniam Sundranar University, Tirunelveli. He has publised research papers on clustering algorithm in international journals.

E-mail: jaypeeyes@rediffmail.com

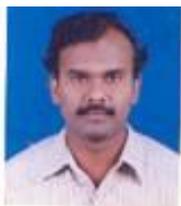 **Dr. S. P. Victor** earned his M.C.A. degree from Bharathidasan University, Tiruchirappalli. The M. S. University, Tirunelveli, awarded him Ph.D. degree in Computer Science for his research in Parallel Algorithms. He is the Head of the department of computer science, and the Director of the computer science research centre, St. Xavier's college (Autonomous), Palayamkottai, Tirunelveli. The M.S. University, Tirunelveli and Bharathiar University, Coimbatore have recognized him as a research guide. He has published research papers in international, national journals and conference proceedings. He has organized Conferences and Seminars at national and state level.
E-mail: victorsp@rediffmail.com